\documentclass[journal]{IEEEtran}
\ifCLASSINFOpdf
\else
\fi

\usepackage{graphicx}
\usepackage{textcomp}
\usepackage{xcolor}
\usepackage{colortbl}
\definecolor{mygray}{gray}{0.98}
\definecolor{mygray1}{gray}{0.93}
\usepackage{algorithm}  
\usepackage{algpseudocode}  
\usepackage{amsmath}  
\usepackage{amsfonts}
\usepackage{amssymb}
\usepackage{latexsym}
\usepackage{caption, subcaption}
\usepackage{enumitem}
\usepackage{booktabs}
\usepackage{listings}
\usepackage{chngpage}
\usepackage{flushend}
\usepackage{adjustbox}
\usepackage{multirow}
\usepackage{textcomp,booktabs}
\newcommand{\circled}[1]{\normalsize{\textcircled{\scriptsize{#1}}}\normalsize\;}
\usepackage{ragged2e}
\usepackage{xcolor}
\usepackage[colorlinks,
            linkcolor=blue,
            anchorcolor=blue,
            citecolor=blue]{hyperref}

\definecolor{seagreen}{rgb}{0.18, 0.55, 0.34}
\definecolor{royalpurple}{rgb}{0.47, 0.32, 0.66}
\definecolor{brown(traditional)}{rgb}{0.59, 0.29, 0.0}
\definecolor{blue}{rgb}{0.3, 0.2, 0.9}

\begin{document}
%
\title{Optimizing Mobile-Edge AI-Generated Everything (AIGX) Services by Prompt Engineering: Fundamental, Framework, and Case Study}
%
%
%

\author{Yinqiu~Liu,
        Hongyang~Du,
        Dusit~Niyato,~\IEEEmembership{Fellow,~IEEE},
        Jiawen~Kang,
        Shuguang~Cui,~\IEEEmembership{Fellow,~IEEE},
        Xuemin~(Sherman)~Shen,~\IEEEmembership{Fellow,~IEEE},
        and
        Ping Zhang,~\IEEEmembership{Fellow,~IEEE}
    \thanks{Y. Liu, H. Du, and D. Niyato are with the School of Computer Science and Engineering, Nanyang Technological University, Singapore (e-mail: yinqiu001@e.ntu.edu.sg, hongyang001@e.ntu.edu.sg, dniyato@ntu.edu.sg). J. Kang is with the School of Automation, Guangdong University of Technology, China (e-mail: kavinkang@gdut.edu.cn).  S. Cui is with the School of Science and Engineering (SSE) and the Future Network of Intelligence Institute (FNii), The Chinese University of Hong Kong (Shenzhen), China (e-mail: shuguangcui@cuhk.edu.cn). X. Shen is with the Department of Electrical and Computer Engineering, University of Waterloo, Canada (e-mail: sshen@uwaterloo.ca). P. Zhang is with the State Key Laboratory of Networking and Switching Technology, Beijing University of Posts and Telecommunications, China (e-mail: pzhang@bupt.edu.cn). \vspace{-1em}}
        }

\maketitle

\begin{abstract}
As the next-generation paradigm for content creation, AI-Generated Content (AIGC), i.e., generating content automatically by Generative AI (GAI) based on user prompts, has gained great attention and success recently.
With the ever-increasing power of GAI, especially the emergence of Pretrained Foundation Models (PFMs) that contain billions of parameters and prompt engineering methods (i.e., finding the best prompts for the given task), the application range of AIGC is rapidly expanding, covering various forms of information for human, systems, and networks, such as network designs, channel coding, and optimization solutions. 
In this article, we present the concept of mobile-edge AI-Generated Everything (AIGX). 
Specifically, we first review the building blocks of AIGX, the evolution from AIGC to AIGX, as well as practical AIGX applications.
Then, we present a unified mobile-edge AIGX framework, which employs edge devices to provide PFM-empowered AIGX services and optimizes such services via prompt engineering.
More importantly, we demonstrate that suboptimal prompts lead to poor generation quality, which adversely affects user satisfaction, edge network performance, and resource utilization. 
Accordingly, we conduct a case study, showcasing how to train an effective prompt optimizer using ChatGPT and investigating how much improvement is possible with prompt engineering in terms of user experience, quality of generation, and network performance.
\end{abstract}


%
\IEEEpeerreviewmaketitle

\section{Introduction}
In 2022, the phenomenal success of DALL$\cdot$E 2 and ChatGPT ignited the research and applications of Generative Artificial Intelligence (GAI).
Different from the traditional discriminative AI that learns the boundaries among classes, GAI can represent complex distributions and generate realistic samples following the same distributions \cite{du2023deep}.
Thanks to sophisticated representation and generation ability, GAI realizes content creation in various modalities, such as texts, images, and videos.
Nowadays, the content generated automatically by GAI based on user-input keywords (i.e., prompts) is called AI-Generated Content (AIGC) \cite{du2023deep}.

Despite the boom in 2022, GAI has been studied for years and experienced several rounds of evolutions.
Particularly, the emergence of Pretrained Foundation Models (PFMs) and prompt-based methods have greatly facilitated the development of GAI and AIGC.
We take Natural Language Processing (NLP) as an example:
By pretraining on huge raw datasets, PFMs, with hundreds of billions of learnable parameters, can first learn massive general-purpose linguistic knowledge, including words, grammar, and sentence composition.
Then, different from fine-tuning, which adjusts the model parameters, prompt-based methods keep PFM frozen and only reformulate the downstream tasks (e.g., text translation and summarization), associating unseen tasks with the pre-learned knowledge.
Circumventing tedious task-oriented fine-tuning, such a flexible ``PFM\,+\, prompt" paradigm significantly expands the application range of GAI beyond generating human-oriented multimedia content. 
As the automatic creations of channel coding \cite{du2023deep}, network designs \cite{zou2023wireless}, and defenses \cite{ferrag2023revolutionizing} are realized successively, AIGC is evolving to the new stage, which we coin as AI-Generated Everything (AIGX).
However, two obstacles still exist from AIGC to AIGX, namely resource constraints and low-quality prompts.

\begin{itemize}
    \item \textbf{Resource Constraints}: Containing numerous parameters, PFMs are resource-intensive. For instance, deploying GPT-3 requires at least one NVIDIA Ampere or even newer GPU with no less than eight GB of GPU memory \cite{liuaigc}. Moreover, each round of generative inference consumes substantial computing power. Such huge resource consumption is unaffordable for many resource-constrained mobile users.
    \item \textbf{Low-Quality Prompts}: Composing professional prompts to instruct the PFMs effectively is challenging for untrained users, especially when the downstream tasks are complicated or the selected PFM has latent requirements. Low-quality prompts may decrease the PFMs' generation quality, even bringing in security concerns \cite{prompt}.
\end{itemize}

\begin{figure*}[tbp]
\centerline{\includegraphics[width=1.93\columnwidth]{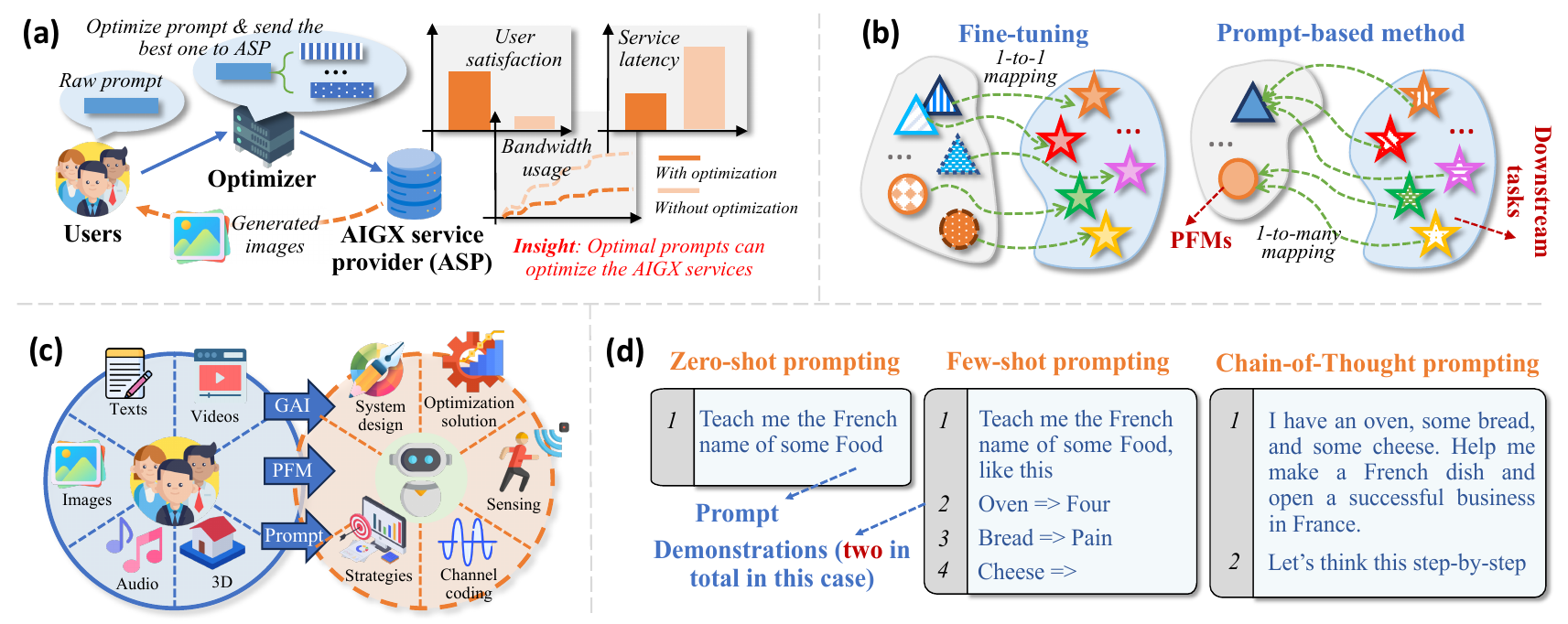}}
\caption{\texttt{(a)} The image generation case in mobile-edge AIGX. \texttt{(b)} The fine-tuning and prompt-based method diagrams. Fine-tuning forms a 1-to-1 mapping, i.e., each PFM can apply only to one task. However, by freezing PFMs, the prompt-based method customizes the tasks (denoted by different textures) and forms the 1-to-many mapping. This is more efficient since no computing resources and datasets for fine-tuning are required. \texttt{(c)} The illustration of the evolution from AIGC to AIGX. \texttt{(d)} The illustration of zero-shot, few-shot, and Chain-of-Thought prompting.}
\label{prompt}
\end{figure*}

These challenges become more critical in mobile environments where computing, communication, and data resources are limited.
Fortunately, edge computing and prompt engineering show great potential to address these issues.
Firstly, with enough physical resources, edge servers can operate PFMs locally and serve as the AIGX Service Providers (ASPs) to perform inferences for mobile users \cite{liuaigc}.
In such a mobile-edge AIGX environment, one of the most fundamental problems is finding the most appropriate prompt for the given tasks and PFMs, so-called \textit{prompt engineering} \cite{prompt}.
Prompt engineering plays a critical role in optimizing mobile-edge AIGX services.
It is worth noting that prompt engineering for mobile-edge AIGX could be regarded as a general network optimization problem, in which prompts, as decision variables, have to be selected and structured to achieve the user objective under a set of network constraints.
Take the route selection problem in vehicular networks as an example, in which the relay nodes must be scheduled to form the optimal path. 
Otherwise, the packet transmission will suffer from high latency and loss due to the limited bandwidth and congestion.
Likewise, Fig. \ref{prompt}(a) illustrates an image generation scenario in mobile-edge AIGX \cite{hao2022optimizing}, where the user employs an optimizer that optimizes the raw prompts for ASPs to the optimal.
The role of prompt engineering includes:
\begin{itemize}
    \item \textbf{Improving Quality of Generation}: The power of PFMs can be maximized only if the optimal prompts are given. According to Hao \textit{et al.} \cite{hao2022optimizing}, optimizing prompts can increase users' satisfaction level towards the generated images by 380\%.
    \item \textbf{Improving Service Experience}: A higher satisfaction level means a smaller number of re-generations required. Consequently, the overall service latency and service fee can be decreased, which leads to the increased Quality of Experience (QoE) from the user perspective.
    \item \textbf{Reducing Energy Consumption}: Resource allocation is critical in mobile-edge networks. Reducing the number of re-generations can reduce the waste of bandwidth and computing power.
\end{itemize}

Motivated by this, we introduce mobile-edge AIGX services in this article, especially its optimization by prompt engineering.
The contributions of this paper are summarized as follows:
\begin{itemize}
    \item \textit{To the best of our knowledge, this is the first work that systemically introduces mobile-edge AIGX.} To begin with, we review the building blocks of AIGX and the evolution from AIGC to AIGX; we then introduce four practical applications of AIGX.
    \item We present a unified framework for mobile-edge AIGX, which leverages edge devices to provide PFM-empowered AIGX services and optimizes services by prompt engineering methods.
    \item We take the VR-empowered interior design as a case study, in which we train a ChatGPT-empowered prompt optimizer and explore how much improvement prompt engineering can achieve regarding QoE, service fee, and bandwidth consumption.
\end{itemize}

\section{AIGX: Building Blocks, Basics, and Applications}

\subsection{GAI \& PFMs}
GAI and PFMs are the major building blocks for AIGX.
Specifically, GAI enables machines to generate realistic samples, e.g., texts, images, videos, and even 3D avatars, following the target distribution.
PFMs support GAI by packing well-trained models that master massive general-purpose knowledge from huge raw datasets, relieving us from training GAI from scratch.
Here, we briefly introduce some representative GAI architectures and the corresponding PFMs.
\begin{itemize}
    \item \textbf{Transformer}: The transformer model revolutionizes NLP by introducing a self-attention mechanism that captures contextual relationships in text. It powers many SOTA PFMs such as GPT-4, RoBERTa, CLIP, and SAM, enabling various tasks in NLP (e.g., translation, summarization, and Q\&A) and computer vision (e.g., text-to-image mapping and image segmentation) \cite{PFMscollection}.
    \item \textbf{GAN}: In Generative Adversarial Network (GAN) model, the generator network attempts to synthesize plausible samples, while the discriminator network keeps learning the detection of synthesized samples from real ones \cite{du2023deep}. The generator network can gradually find an optimal generation strategy through the game with the discriminator network. GAN sparked many advancements in image processing, such as StyleGAN for image synthesis and style transfer \cite{PFMscollection}.
    \item \textbf{Diffusion}: Inspired by nonequilibrium thermodynamics, the diffusion model generates content by denoising from Gaussian noise \cite{du2023deep}. Such a denoising process is guided by the inverse process, i.e., gradually perturbing real-world samples, which is pre-trained. Diffusion powers many famous PFMs, such as DALL$\cdot$E 2, Midjourney, and Stable Diffusion in text-to-image generation, Make-A-Video in video generation, and DreamFusion in 3D generation \cite{PFMscollection}. 
\end{itemize}

\subsection{Bridging PFMs and Tasks by Prompts}
As shown in Fig. \ref{prompt}(b), prompts can be regarded as the reformulation of downstream tasks, making unseen tasks look more similar to those solved during the pretraining.
Considering the heterogeneous modalities and implementation of PFMs, the prompts can take various forms, including:
\begin{itemize}
    \item \textbf{Textual Prompts}: Textual prompts refer to the textual description of the downstream tasks. For instance, the user-input keywords for guiding Stable Diffusion and the questions we ask ChatGPT are textual prompts.
    \item \textbf{Visual Prompts}: Visual prompts are adopted in visual and visual-language PFMs. For instance, users' clicks and bounding boxes for guiding the semantic segmentation in SAM are visual prompts. 
    \item \textbf{Multimodal Prompts}: Prompts in multiple modalities can be fused to improve the task representation or be fed to different modules of multimodal PFMs \cite{Khattak_2023_CVPR}, supporting multimodal tasks such as visual-language mapping.
    \item \textbf{Continuous Prompts}: Prompts are for task reformulation rather than for human interaction. Therefore, they are not necessary to be limited to human-interpretable texts or images and can be parameterized tokens or layers appended to PFMs, e.g., the embedding vectors for preprocessing input images in VPT \cite{VPT}. 
\end{itemize}
Prompt-based method refers to using prompts to bridge PFMs with diverse downstream tasks and is viewed as the key to unleashing the power of PFMs. As shown in Fig. \ref{prompt}(b), instead of loading the entire PFMs to memory and fine-tuning the model parameters on task-specific datasets, prompt-based methods only reformulate downstream tasks with customized prompts while keeping PFM frozen, achieving the following advantages.
\begin{itemize}
    \item \textbf{Few-Shot Learning Enablement}: Prompt-based method enables PFMs to solve downstream tasks by learning from only a few labeled demonstrations, making it an efficient strategy when available demonstrations are scarce.
    \item \textbf{High Efficiency}: Considering the large size, operations on PFMs are time-consuming and resource-intensive \cite{prompt}. In contrast, keeping the PFM frozen and only adjusting the inputs, i.e., textual, visual, or parameterized prompts, is much less expensive regarding resources and time.  
    \item \textbf{Versatility}: Different from fine-tuning, which is both task-oriented and model-specific, prompt-based methods only reformulate downstream tasks, enabling one PFM to support various tasks [see Fig. \ref{prompt}(b)]. Such a feature can greatly expand the application ranges of PFMs.
\end{itemize}


\subsection{From AIGC to AIGX}
As shown in Fig. \ref{prompt}(c), the initial scope of AIGC is limited to generating human-oriented multimedia content, such as articles, music, and videos.
Nonetheless, the application range of AIGC is rapidly expanding due to the momentum from three aspects. 
First, with the advancements in GAI models, more complicated types of information can be generated. For instance, researchers have successfully leveraged diffusion models to generate channel coding, robot trajectories, etc.
Moreover, the hardware evolution provides us with enough computing resources to train PFMs with massive knowledge in different subjects and powerful capabilities to understand the context.
Hence, more complex prompts and environment descriptions can be applied to PFMs, which is a prerequisite for PFMs to accomplish complex generation tasks.
Finally, the prompt-based methods greatly expand the applicability of PFMs, allowing one PFM to support diverse downstream applications.
Nowadays, a series of information beyond multimedia content can be generated by GAI, e.g., network designs \cite{zou2023wireless} and cybersecurity defenses \cite{ferrag2023revolutionizing}.
Therefore, we present the concept of \textit{AI-Generated Everything}, which covers all possible forms of information generated by GAI for humans, machines, and systems.
Next, we show several examples of AIGX.

\subsection{Examples of AIGX}
\subsubsection{AI-Generated System Design} 
PFMs show great potential in scheduling complex networks and systems due to their analytical and reasoning abilities.
For instance, Zou \textit{et al.} \cite{zou2023wireless} employ ChatGPT to schedule the transmission power of senders for saving energy in shared-spectrum communications.
By informing ChatGPT the \textit{environment} (i.e., $n$ senders, $m$ base stations, $r$ rounds, bandwidth $b$ kHz, channel gain $g$, noise $n$ dBm/Hz), \textit{game rules} (i.e., in each round, each sender chooses its transmission power), and \textit{network goal} (i.e., the total power consumption does not exceed $p$ W), a qualified transmission power schedule of each sender in any round can be generated automatically.  

\subsubsection{AI-Generated Data Processing} 
For efficiency or privacy purposes, the message for transmission can be processed generatively, i.e., using GAI to generate compressed or privacy-preserving contents from raw data. 
For example, Liu \textit{et al.} \cite{liu2023semantic} present a generative semantic extraction method for compressing data transmitted over wireless links.
Take image transmission as an example.
Unlike traditional semantic communications that extract features from raw images, by ControlNet \cite{PFMscollection}, the sender directly generates raw images' skeletons, which are transmitted to the receiver and decoded by a Stable Diffusion.
Due to data size compression, the bandwidth consumption is reduced by half while most semantic features are preserved.

\subsubsection{AI-Generated Optimization Solution} 
Optimization widely exists in practical system applications as a problem of finding the best solution from all feasible ones. 
Recently, diffusion models showed great potential in optimization since the denoising process starting from Gaussian noise can be regarded as the optimization process of finding the optimal solution starting from random policies.
In light of this, Du \textit{et al.} \cite{du2023deep} introduce a diffusion process to Deep Reinforcement Learning (DRL), which greatly improves the exploration ability and achieves the SOTA performance in many optimization problems, such as incentive mechanism design, wireless resource allocation, and channel coding for wireless communications.

\subsubsection{AI-Generated Programming}
Besides generating human-oriented texts, GAI also supports generating machine-oriented codes in different languages. 
For instance, Destefanis \textit{et al.} \cite{destefanis2023preliminary} utilized GPT-3.5 to generate Java codes and proved that AI-generated codes can solve 90.5\% of practice coding problems published on \textit{CodingBat.com} website.
Furthermore, code translation, e.g., rewriting MATLAB codes by Python, and code documentation, i.e., generating user manuals for source codes, can also be realized by GAI.

\textbf{Lesson Learned}:
The above examples demonstrate that AIGX has the following features:
\begin{itemize}
    \item \textbf{Interaction with Environment}: Apart from serving human users, in many cases, the ``things" generated by AIGX should interact with the environment, such as the AI-generated smart contracts for blockchain systems. 
    \item \textbf{Sensitivity to Uncertainty}: AIGX is less tolerant of unexpected outcomes. In the above example, if the generated code contains loopholes, the digital assets will be subject to a great risk of loss.
    \item \textbf{Requirement for Efficiency}: AIGX not only concerns the user-level generation quality but also focuses on improving the system-level performance, such as latency, resource efficiency, robustness, etc.
\end{itemize}
These features put forward urgent requirements for high-quality prompts, only by which the environments can be described clearly, and unexpected generation errors can be alleviated. Moreover, we intend to explore the role of prompt engineering in optimizing not only the user experience but also the efficiency of the AIGX services.

\section{Optimizing Mobile-Edge AIGX by Prompt Engineering}
In this section, we demonstrate the optimization of mobile-edge AIGX by prompt engineering, including the fundamentals of prompt engineering and the unified framework that incorporates prompt engineering in mobile-edge AIGX. 

\subsection{Prompt Engineering Methods}
\subsubsection{Manual Methods}
A typical way of prompt engineering is to compose different prompts and search for the one that leads to the best generation quality manually.
\begin{itemize}
    \item \textbf{Content Optimization}: Firstly, we can refine the prompt content, thereby better associating the downstream tasks with the learned knowledge. For example, ChatGPT prompts can be refined by clarifying the \textit{role} (i.e., what role that ChatGPT is expected to act as, e.g., a writer), the \textit{objective} (i.e., what is the purpose of this prompt, e.g., comparing fine-tuning and prompt-based methods), and the \textit{outcome} (i.e., what is the format of the expected outcome, e.g., a table showing the difference)\footnote{https://the-decoder.com/chatgpt-guide-prompt-strategies/}. 
    \item \textbf{Organization Optimization}: The prompting strategy also greatly affects the generation quality. For instance, Fig. \ref{prompt}(d) shows three organizations of textual prompts, namely zero-shot prompting, few-shot prompting, and Chain-of-Thought (CoT) prompting \cite{fewshot}. Firstly, zero-shot prompting means that only a task description is provided. In contrast, few-shot prompting gives PFMs a few demonstrations to showcase the expected output format. Finally, for some deductive tasks, the CoT paradigm can help PFMs split the task into multiple simple subtasks and generate correct results step by step.
\end{itemize}

\subsubsection{Automatic Methods}
Although manual methods are easily accessible, they require the users' prompting experience and an in-depth understanding of the downstream tasks.
Therefore, researchers further present various automatic prompt engineering methods, circumventing the effects caused by human-related factors, e.g., limited knowledge and bias, and achieving stable prompt optimization performance.
\begin{itemize}
    \item \textbf{Discrete Prompts}: Discrete prompts, such as texts and images, are typically human-interpretable. Take the optimization of the textual prompts as an example. Jiang \textit{et al.} \cite{jiang-etal-2020-know} presented the prompt paraphrasing. Specifically, each given raw prompt is translated into multiple languages, during which some phrases were replaced by those from a thesaurus. In this way, a series of prompts can be generated, which are then sent to PFMs to achieve the highest generation quality. Moreover, thanks to the evolution of PFMs, the optimized prompts can even be directly generated \cite{prompt}.
    \item \textbf{Continuous Prompts}: In contrast to discrete prompts, continuous prompts mainly refer to the learnable embedding layer, which pre-processes the inputs before feeding them to PFMs. Consequently, such prompts can be refined by fine-tuning their learnable parameters, called prompt tuning \cite{prompt}. 
\end{itemize}
\begin{figure*}[tbp]
\centerline{\includegraphics[width=1.82\columnwidth]{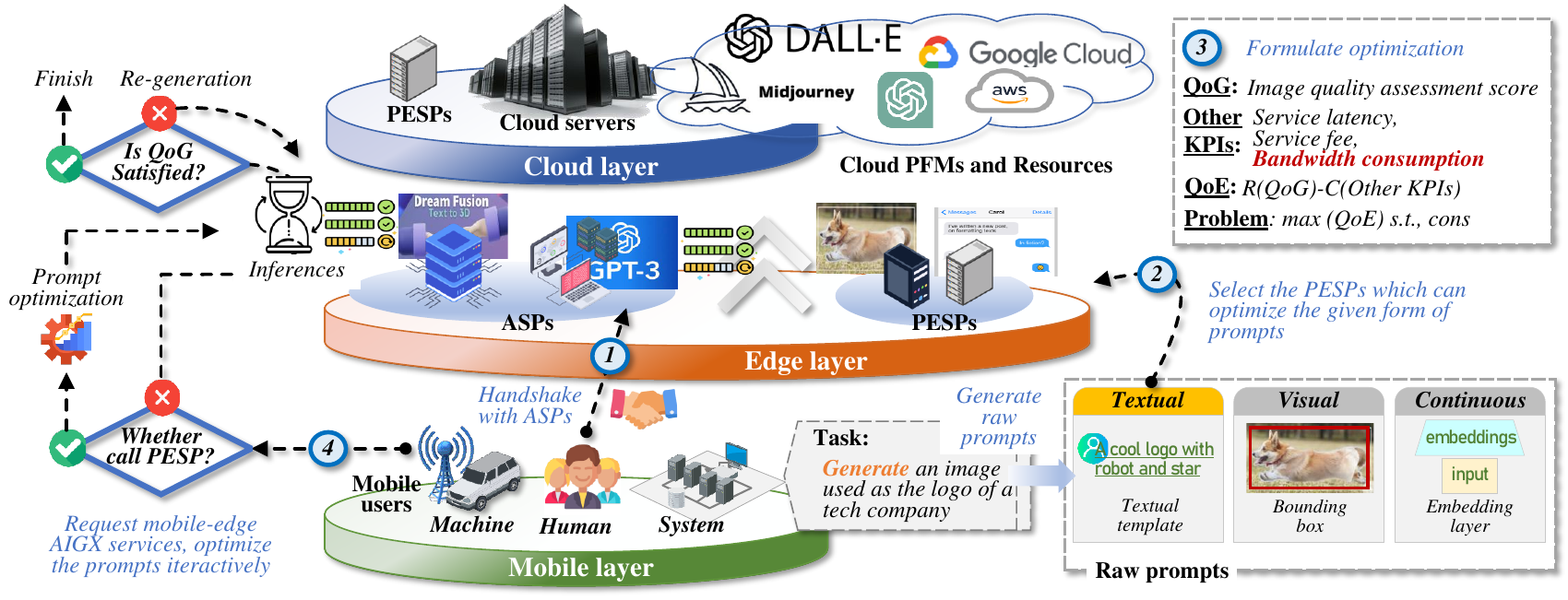}}
\caption{The framework design. The three-layer framework architecture is shown in the middle. In addition, we take the AI-generated image application as an example to illustrate the four steps that users should perform.}
\label{frame}
\end{figure*}

\subsection{Framework Design}
In this part, we present a unified mobile-edge AIGX framework, which provides PFM-empowered AIGX services and adopts prompt engineering for optimizing the service.

\subsubsection{Architecture}
As shown in Fig. 2, our framework follows a mobile-edge-cloud architecture like NetGPT \cite{chen2023netgpt}.
The mobile layer accommodates mobile users, who request AIGX services and use the generated ``things" to empower various applications, such as Web 3.0 and Metaverse.
It is worth noticing that ``users" can refer to not only humans who require multimedia content but also machines and networks that require system designs, codes, etc.
However, users suffer from limited physical resources.
Above the mobile layer, numerous edge servers operate PFMs locally and serve as ASPs in the so-called edge layer.
The introduction of an edge layer can bring the following advantages.
First, the data transmissions between these two layers are realized by mobile-edge communications. 
Since ASPs are close to users, low communication latency can be guaranteed. 
Moreover, users can access nearby ASPs and receive the ubiquitous and customized AIGX services.
Finally, the system's robustness can be enhanced because the introduction of massive ASPs can eliminate single-point failures and also avoid congestion on a single server.
The cloud layer supports the lower layers by 1) providing storage services, e.g., the datasets for training PFMs, and 2) operating extremely large PFMs, such as GPT-4 and DALL$\cdot$E 2, with high performance.
Accordingly, users are free to call cloud-based ASPs with more powerful PFMs if edge-based ASPs cannot meet their demands.
To realize prompt engineering in our framework, Prompt Engineering Service Providers (PESPs) are deployed in the edge and cloud layers.
In this case, users can send raw prompts to PESPs and request paid prompt engineering services.
Similar to ASPs, users can choose edge- or cloud-based PESPs according to the specific application requirements. Compared with edge-based PESPs, cloud-based PESPs are more powerful since the cloud servers have enough computation and storage resources to support extremely large datasets and machine learning models for training complicated prompt engineering methods. However, since massive users send prompts to one cloud server simultaneously, cloud-based prompt engineering may suffer from long service latency and leakage of user privacy.
\begin{figure*}[tbp]
\centerline{\includegraphics[width=1.95\columnwidth]{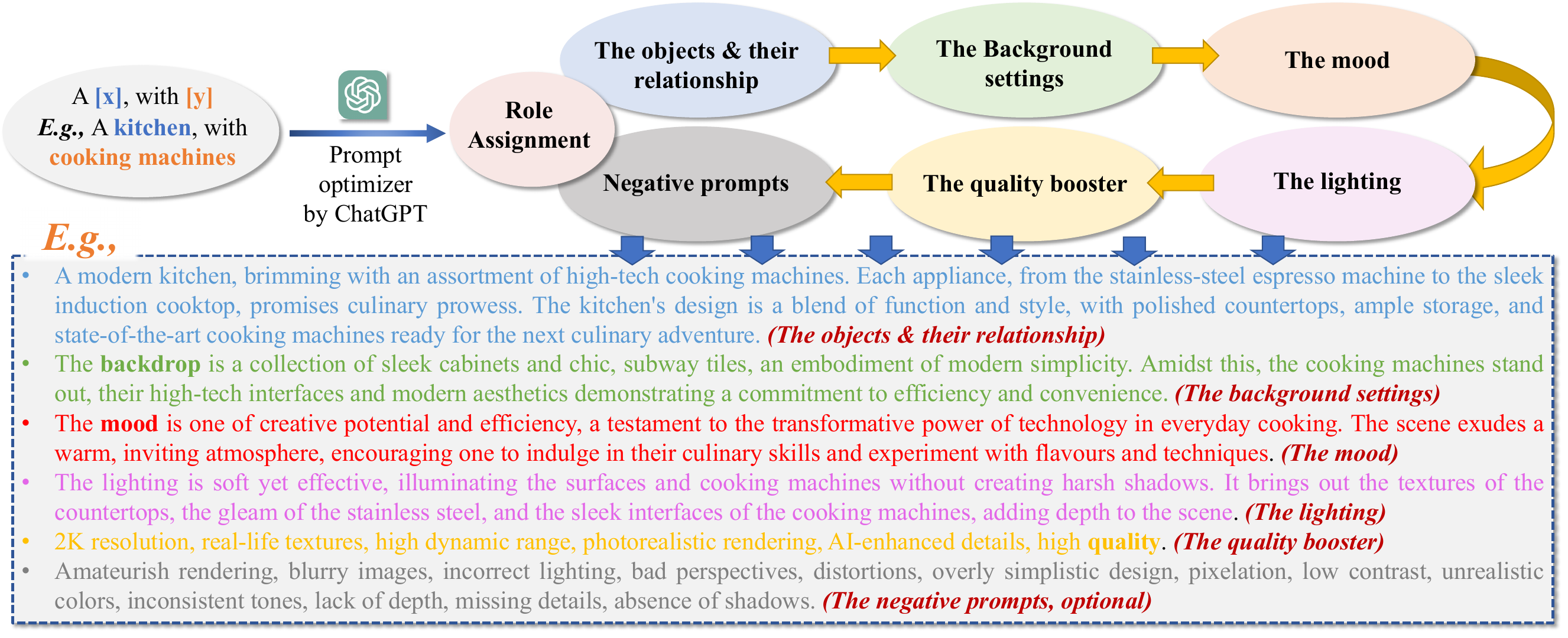}}
\caption{The ChatGPT-empowered prompt optimizer and an example about the kitchen design. We can observe that the entire prompt template consists of six successive parts.}
\label{prompt3}
\end{figure*}

\subsubsection{User Workflow}
In the proposed framework, to request mobile-edge AIGX services and acquire optimal performance, users should perform the following four steps (see Fig. \ref{frame}).
\begin{itemize}
   \item \textbf{Service Configuration}: Considering the diversity of PFMs, the ASPs are highly heterogeneous in terms of the AIGX services that they can provide. Therefore, the first step for users is to find suitable ASPs whose PFMs' capability can meet their specific requirements. The users can conduct multiple rounds of handshakes with target ASPs, confirming the form of the required information (e.g., an image, an article, or an optimization solution), expected service fee, and other configurations. Take AI-generation images as an example. The ASP should run text-to-image models and be able to generate images with different styles.
    \item \textbf{Prompt Engineering Configuration}: Similar to traditional optimization problems, in which the optimization strategies should be designed according to the specific types of optimization variables, different types of prompts, e.g., textual, visual, or parameterized prompts, also correspond to different prompt engineering methods. Therefore, in this step, the users select suitable PESPs to optimize a given form of prompts. In our example, the selected PESP should be trained on massive image-caption pairs and linguistic corpus, thereby improving the precision of the raw prompt in describing the image.
    \item \textbf{Optimization Formulation}: Then, the users optimize the AIGX services. Specifically, a series of Key Performance Indicators (KPIs) can be considered to define the objective functions. At the user level, the most fundamental KPI is the \textit{Quality of Generation (QoG)}. According to the specific AIGX tasks, QoG can be defined in different ways. For instance, in our example, QoG can be measured by image quality assessment scores \cite{talebi2018nima}. Higher QoG not only represents higher user satisfaction levels but also reduces the service latency, service fee, and system-level bandwidth consumption caused by re-generations. However, calling PESPs causes extra service fees and bandwidth consumption. Facing this trade-off, the users should define their own QoG and build personalized QoE models. As shown in Fig. \ref{frame}-Step \circled{3}, a simple QoE model can take the form of \textit{reward} - \textit{cost}, in which the \textit{reward} is determined by QoG and the \textit{cost} is related to service latency, service fee, bandwidth consumption, etc. Accordingly, the optimization objective is to maximize the QoE. Moreover, the constraints from users and environments, e.g., the available service fee and resources of users and PESPs, respectively, should be considered. 
    \item \textbf{AIGX Inference \& Iterative Refinement}: Finally, the users call AIGX services. Specifically, they can improve the QoE by adjusting the policy for conducting prompt engineering. As shown in Fig. \ref{frame}-Step \circled{4}, in each round, the users decide whether to call PESP and measure the resulting QoG. If the value satisfies the users' requirements, they can enter the next round. Otherwise, they ask ASPs for re-generation. Note that the effectiveness of the selected prompt engineering method in a given task is a posteriori knowledge to the users. Hence, they have to monitor their long-term QoE and choose more powerful PESPs when the current PESP cannot increase QoE anymore.  
\end{itemize} 

\section{Case Study: Resource-efficient VR-Based Interior Design via Prompt Optimization}
In this section, we conduct a case study to showcase the optimization of mobile-edge AIGX service by prompt engineering and utilize numerical results to demonstrate how prompt engineering can help improve AIGX efficiency.

\subsection{System Model}
We consider a VR creation scenario where designers need to come up with different interior design solutions to serve customers. 
To do so, the designers keep generating plentiful images depicting different interior scenes with diverse styles and use them as the materials for rendering the virtual rooms\footnote{The image-3D rendering can be realized by many tools, such as NVIDIA instant NeRF: https://github.com/NVlabs/instant-ngp}.
Following the proposed framework, the designers will offload the image generation tasks to ASPs, which use mainstream text-to-image models \cite{PFMscollection} to realize the text-to-image AIGX.
PESPs are also implemented using the ChatGPT-empowered prompt optimizer proposed below.
\begin{figure*}[tbp]
\centerline{\includegraphics[width=1.85\columnwidth]{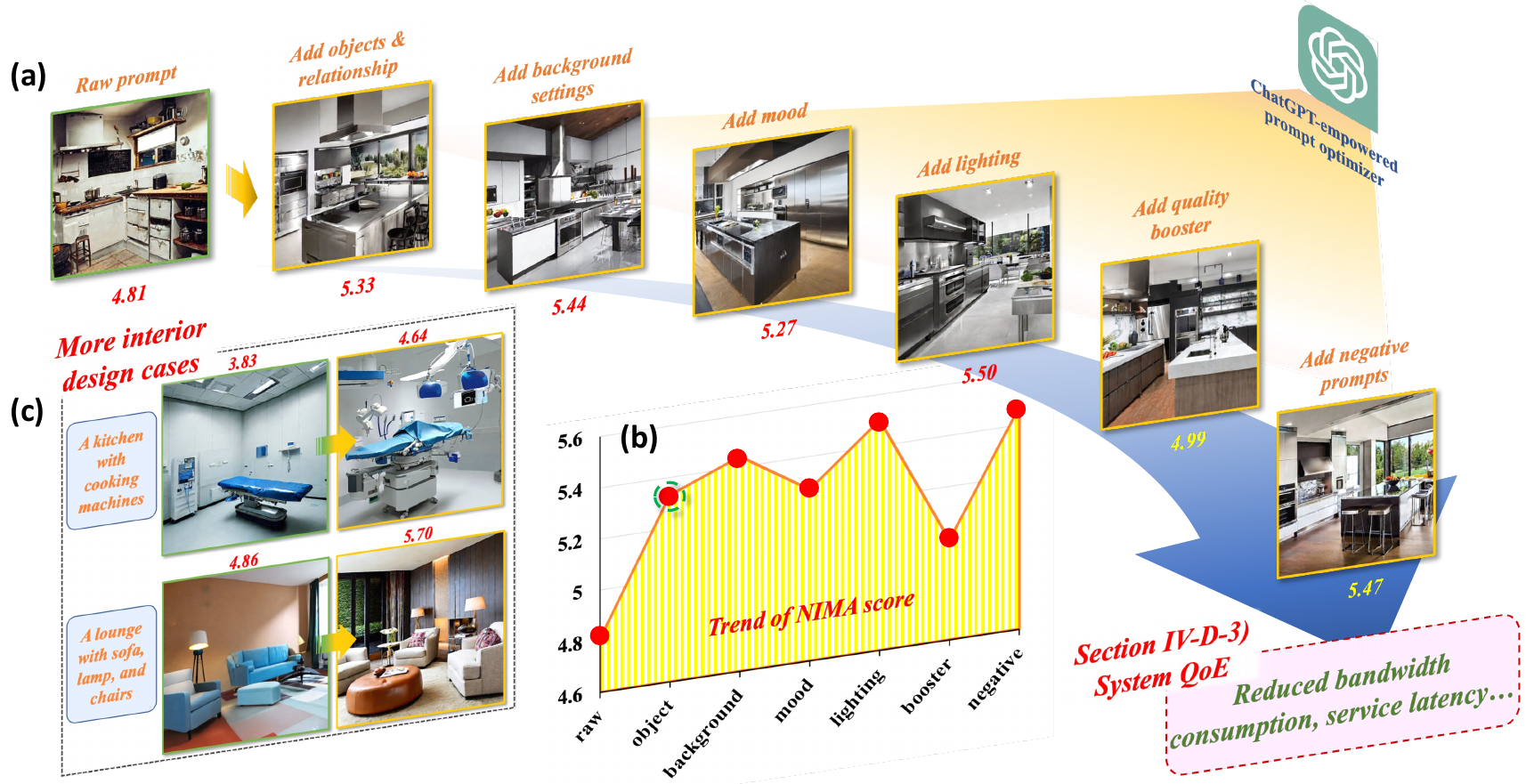}}
\vspace{0.1cm}
\caption{The effectiveness of ChatGPT-empowered prompt optimizer. \texttt{(a)} The images of ``\textit{A kitchen, with cooking machines}" using the raw and optimized prompts. Note that we perform an ablation study to explore which aspect of content enrichment plays an important role in improving QoG. \textit{(b)} The trend of NIMA score in the ablation study. \texttt{(c)} the illustration of other cases for interior design.}
\label{prompt4}
\end{figure*}

\subsection{ChatGPT-Empowered PESP}
In this part, we illustrate how to train a simple but effective prompt optimizer using ChatGPT.
We suppose users generate images using the prompts ``\textit{A [a], with [b]}", e.g., \textit{A kitchen, with cooking machines} for text-to-image generation (see Fig. \ref{prompt3}).
Terms \textit{[a]} and \textit{[b]} represent the macro scene and the iconic objects in it, respectively.
In this way, we can examine the quality of the generated images from the construction of large scenes and the depiction of details. 
We intend to refine raw prompts from the following six aspects\footnote{Inspired by https://www.youtube.com/watch?v=oxW14OBuA8M}.
\begin{itemize}
    \item \textbf{Objects \& Their Relationship}: The raw prompt mentions two types of objects, i.e., a kitchen and cooking machines. The features of objects should be described (e.g., the material and function of the cooking machines). Moreover, to ensure composition correctness, the positional relationship of the objects should be clarified (e.g., how these cooking machines are placed in the kitchen). 
    \item \textbf{Background Settings}: The background settings are used to fill the details for \textit{[a]}. In our example, we add \textit{sleek cabinets} to \textit{[a]}. Such details can enhance the indicative ability of the raw prompts and make the generated image look fine-grained. 
    \item \textbf{Mood}: Mood describes the emotion the creator wants to convey through the image, such as joy or hesitation, which affects the image's tone.
    \item \textbf{Lighting}: Lighting is a crucial factor in determining the texture and authenticity of the AI-generated image. The prompt of lighting includes the light source and the effect of light shining on different objects.
    \item \textbf{Quality Booster}: Quality boosters are adjectives (e.g., \textit{high-quality, 2k resolution}, or \textit{real-life texture}) that facilitate PFMs to recall high-quality images learned at the pretraining stage, thus improving the aesthetic qualities of the generated image. 
    \item \textbf{Negative Prompt}: Negative prompts depict the situations that might decrease the image quality. Note that the terms of negative prompts can be general, e.g., \textit{blurry images} or scene-specific, e.g., \textit{rectangle swimming rings}.
\end{itemize}
Then, we leverage ChatGPT to train an automatic prompt optimizer, which receives raw prompts and atomically enriches their content from these six aspects.
Following the few-shot prompting strategy, we first perform \textit{role assignment}, i.e., asking ChatGPT to act as a professional writer.
Then, we let ChatGPT reformulate the prompt template from ``\textit{A [a], with [b]}" to ``\textit{[1], [2]}, $\dots$, \textit{[6]}", each of which corresponds to one aspect.
We further explain the meaning and the form (e.g., a series of adjectives or concise sentences) of each aspect of ChatGPT.
Finally, we use the example in Fig. \ref{prompt3} as a demonstration for ChatGPT, helping it follow our expected output format. 
The above code is available at: \textit{https://github.com/Lancelot1998/Prompt-Engineering}.

\subsection{Optimization Problem Formulation}
We use NIMA \cite{talebi2018nima} to measure the designer-perceivable QoG towards the generated images.
NIMA is a learning-based framework for image quality assessment, which utilizes neural networks to score the input images and is pre-trained on large aesthetic datasets.
We suppose that the designers only accept images with NIMA scores larger than five. 
Otherwise, they can request for ASPs to re-generate images for free.

Different from AIGC, which only considers QoG, as mentioned in Section II-B, we adopt four KPIs to model the designer-perceivable QoE, namely QoG, service latency, service fee, and the bandwidth consumption of the mobile-edge network.
The rewards are a weight $\alpha$ times the QoG value.
The costs are calculated by fusing the remaining three KPIs linearly, with the weights $\beta_1$, $\beta_2$, and $\beta_3$, respectively. 
In our experiments, we set $\alpha$, $\beta_1$, $\beta_2$, and $\beta_3$ as 15000, 5, 0.5, and 3, respectively.
Finally, we set the service fees for ASPs and PESPs as 10 and 80 monetary units, respectively.

\subsection{Numerical Results}
\subsubsection{Experimental Setup}
Our testbed is a MacBook Pro with an 8-Core Intel Core i9 CPU and AMD Radeon Pro 5500M GPU.
The text-to-image services of ASPs are supported by Stable Diffusion v2.1 \cite{PFMscollection}.

\subsubsection{Aesthetics Analysis}
First, we evaluate the effectiveness of the proposed prompt optimizer in terms of improving the QoG.
As shown in Fig. \ref{prompt4}(a), the image generated by the raw prompt is blurred and poorly composed, with a NIMA score of 4.8.
After prompt optimization, the image quality is significantly increased with modern cooking machines, natural lighting, and higher resolution.
Accordingly, the QoG, i.e., the NIMA score of the image, increases to 5.47.
Moreover, we perform an ablation study, in which we gradually enrich the prompt content, each time adding one more aspect.
Figs. \ref{prompt4}(a) and (b) clearly show that, in this case, optimizing the raw prompt in terms of \textit{objects \& relationships} plays an important role in improving QoG.
However, as mentioned before, the AIGX inferences contain uncertainty.
In other cases, the optimization in other aspects may play a major role.
To this end, as shown in Fig. \ref{prompt4}(c), we further evaluate many other interior design cases, such as the living room and operation room. 
We can conclude that, in general, using the proposed GPT-empowered prompt optimizer can improve the image QoG by an average of 11.85\% and up to 20.5\%.
\begin{figure}[tbp]
\centerline{\includegraphics[width=0.8\columnwidth]{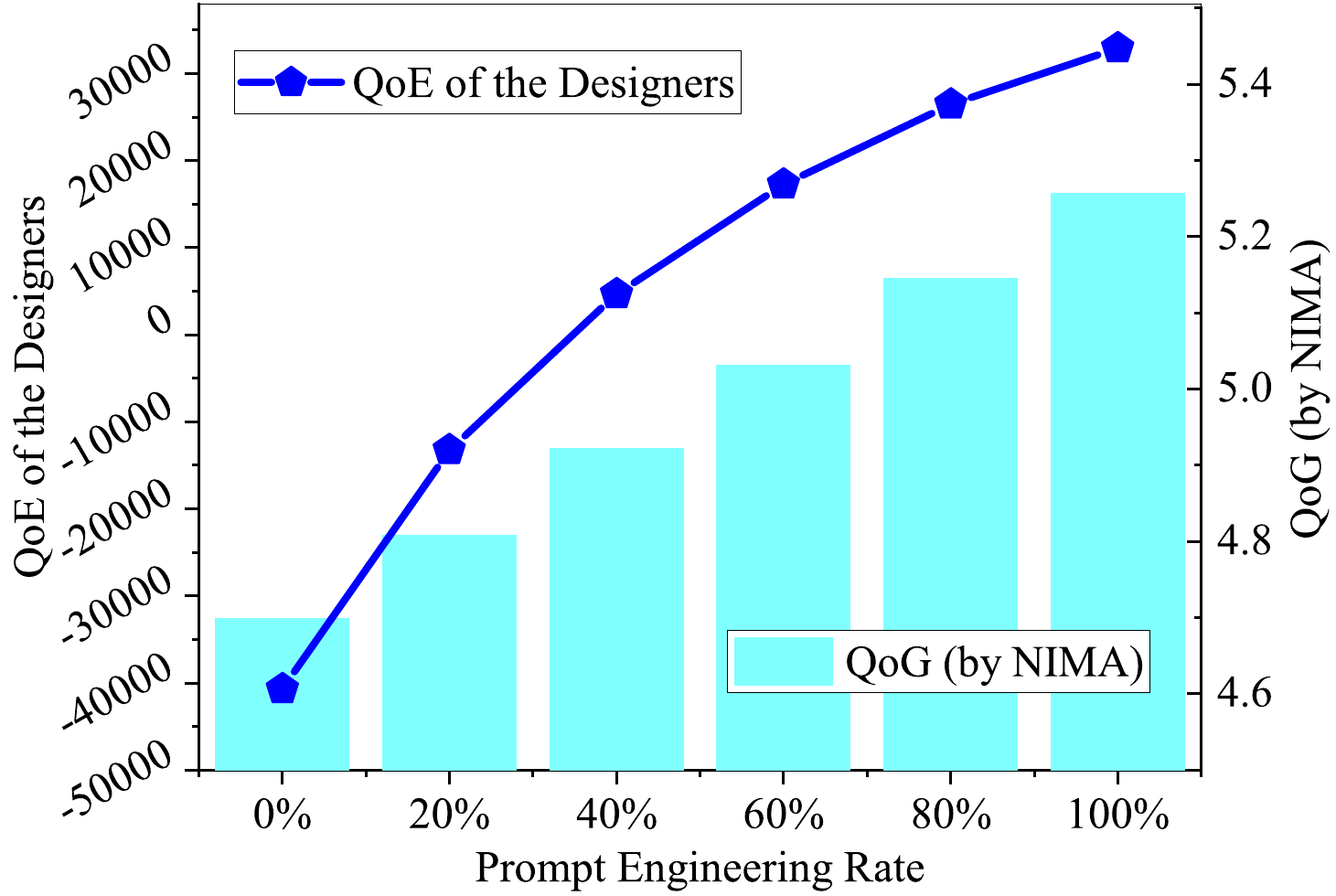}}
\caption{The overall QoE and QoG with different PERs.}
\label{prompt5}
\vspace{0.5cm}
\centerline{\includegraphics[width=0.82\columnwidth]{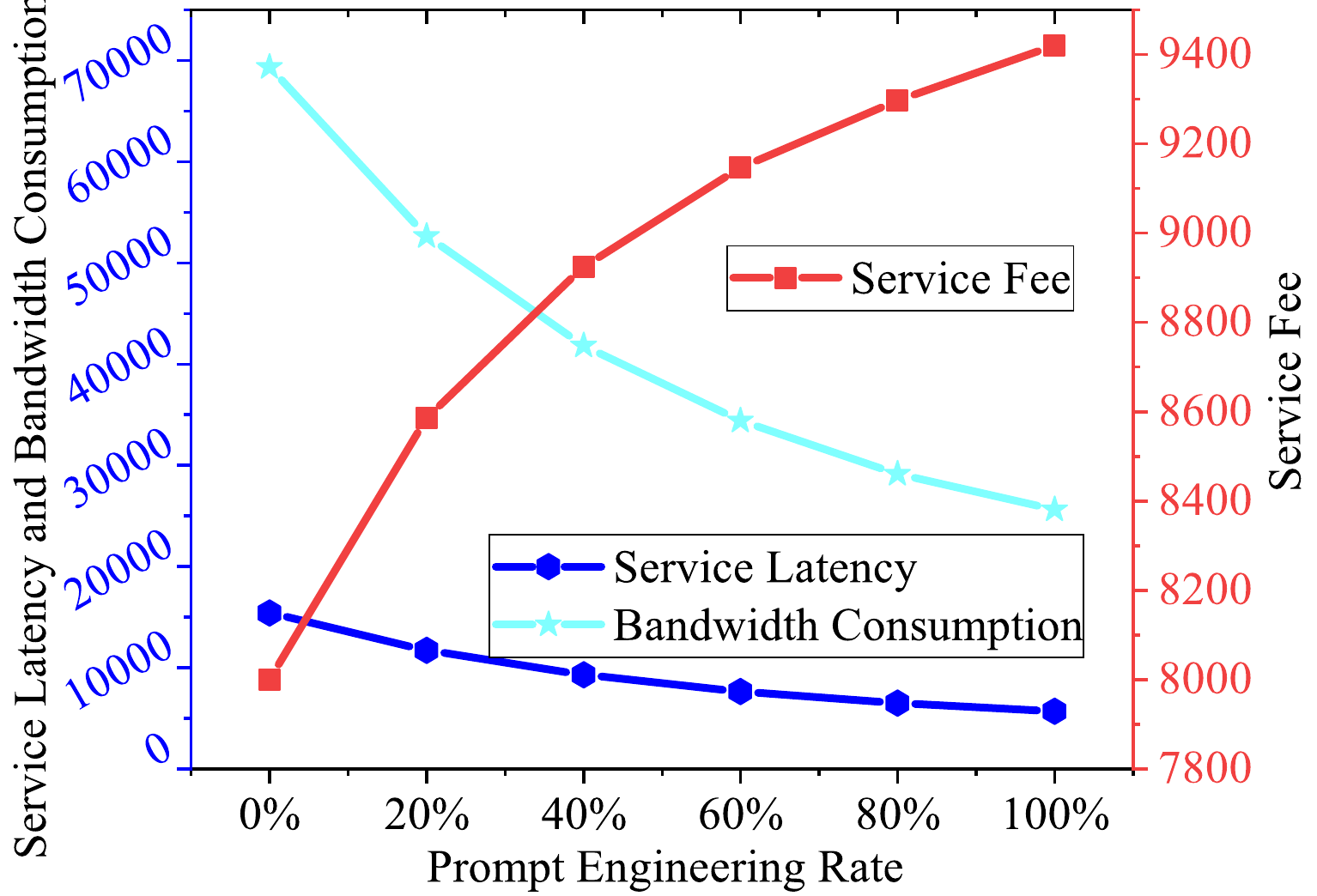}}
\caption{The service latency, service fee, and bandwidth consumption with different PERs.}
\label{prompt6}
\end{figure}

\subsubsection{Numerical Results}
Then, we explore how prompt engineering optimizes the mobile-edge AIGX services.
Fig. \ref{prompt5} shows the QoE of designers for generating 100 images with different prompt engineering rates, i.e., the probability of calling PESPs.
We can observe that the higher the rate, the higher the QoG and the overall QoE.
Without prompt engineering, the QoG of AI-generated images is only 4.698$\pm$0.471, in which case the designers have to re-generate 284 times to acquire 100 qualified images (see Fig. \ref{prompt5}). 
As shown in Fig. \ref{prompt6}, re-generations waste considerable time and network traffic (e.g., 15,402 seconds and 69,310 KB when the prompt engineering rate is 0\%), greatly decreasing the QoE.
With the increasing frequency of calling PESPs, although the service fee increases, the smaller number of re-generations significantly saves time and bandwidth.
In our system, the designers can maximize their QoE by setting prompt engineering rate to 100\%.
Supported by prompt engineering, 100 qualified images can be generated in 5,692 seconds, with only 25,616 KB bandwidth consumption and 43 times of re-generations.

\textbf{Dissusion}: The above experimental results indicate that QoE can be maximized by always performing prompt engineering. 
However, such a conclusion ignores the network constraints.
For instance, if the capability of PESPs is limited and optimizing each prompt takes a long time, the costs for prompt engineering will increase accordingly.
Moreover, as we mentioned in Section II-B, the effectiveness of prompt engineering varies in different tasks.
In cases where optimized prompts cannot increase QoG, calling PESPs will degrade the overall QoE due to the extra service fee. 


\section{Future Directions}

\subsection{Detection of Prompt-Oriented Attacks}
As the reformulation of downstream tasks provided by users, prompts are sensitive and vulnerable to various threats in open mobile-edge networks, such as prompt injection and poisoning. For instance, by appending a specific suffix to ChatGPT prompts, the firewall can be bypassed, and some unhealthy content might be generated. Hence, how to leverage prompt engineering to protect the security of mobile-edge AIGX services, e.g., detecting and filtering abnormal content in the raw prompts, is worth studying.

\subsection{Collaborative Prompt Management in Edge Networks}
In the PFM-empowered AIGX era, prompts are viewed as a novel resource type since only optimal prompts can lead to high QoG, as we demonstrated in this article.
Therefore, to overcome the resource and data limitations of each user, collaborative training and management frameworks for prompts are urgently required.
For instance, researchers have been exploring prompt-oriented federated learning, where users contribute certain datasets and resources to fine-tune the prompts collaboratively.
Moreover, secure prompt sharing or trading can be realized on edge blockchains.

\subsection{Multimodal Mobile-Edge AIGX}
The maturity of AIGX incurs more and more multimodal tasks.
However, a series of problems are yet to be addressed.
Firstly, multimodal prompts are complicated and hard to optimize since the optimization objective should consider the prompt performance in different modalities.
Moreover, defining the rationale QoG for multimodal mobile-edge AIGX services is also intractable and worth research.

\section{Conclusion}
In this article, we presented the concept of mobile-edge \textit{AI-Generated Everything (AIGX)} and discussed its optimization using prompt engineering.
Specifically, we started by introducing the building blocks of AIGX, the evolution from AIGC to AIGX, and real-world AIGX applications.
Then, we reviewed prompt engineering methods and presented a unified framework that incorporates prompt engineering in PFM-empowered mobile-edge AIGX services from an optimization perspective.
Finally, we performed a case study, in which we illustrated how much improvement prompt engineering can lead to in terms of QoE, service latency, and bandwidth usage by numerical results.
As the first article to discuss AIGX, we hope that this article could inspire researchers to explore more applications of AIGX.

\ifCLASSOPTIONcaptionsoff
  \newpage
\fi

\bibliographystyle{IEEEtran}

\vfill

\end{document}